\documentclass[pdflatex,sn-mathphys-num]{sn-jnl}

\usepackage{graphicx}%
\usepackage{multirow}%
\usepackage{amsmath,amssymb,amsfonts}%
\usepackage{amsthm}%
\usepackage{mathrsfs}%
\usepackage[title]{appendix}%
\usepackage{xcolor}%
\usepackage{textcomp}%
\usepackage{manyfoot}%
\usepackage{booktabs}%
\usepackage{algorithm}%
\usepackage{algorithmicx}%
\usepackage{algpseudocode}%
\usepackage{listings}%

\theoremstyle{thmstyleone}%
\theoremstyle{thmstyletwo}%

\theoremstyle{thmstylethree}%

\begin{document}

\title[Article Title]{How Can Machine Learning Accelerate CALPHAD Free Energy Modeling?}


\author*[1]{\fnm{Chen} \sur{Shen}}\email{cshen89@wisc.edu}

\author[1]{\fnm{Muhammad Waqas} \sur{Qureshi}}

\author[2]{\fnm{Mark} \sur{Asta}}

\author[1]{\fnm{Izabela} \sur{Szlufarska}}

\author*[1]{\fnm{Dane} \sur{Morgan}}\email{ddmorgan@wisc.edu}

\affil[1]{\orgdiv{Materials Science and Engineering}, \orgname{University of Wisconsin-Madison}, \orgaddress{\city{Madison}, \postcode{53706}, \state{Wisconsin}, \country{United States}}}

\affil[2]{\orgdiv{Materials Science and Engineering}, \orgname{University of California}, \orgaddress{ \city{Berkeley}, \postcode{94720}, \state{California}, \country{United States}}}


\abstract{The CALPHAD framework provides a rigorous basis for thermodynamic modeling, yet its ability to predict new chemistries is restricted by limited data and by functional forms that rely heavily on composition alone. Here, we show that machine learning (ML) can address these challenges through a hybrid strategy that learns Redlich–Kister (RK) interaction coefficients directly from physically informed elemental descriptors. Using formation energies of 14-element FCC alloys generated by a universal machine-learning interatomic potential (MLIP), we benchmark three classes of models: (i) composition-based RK and ML models, (ii) descriptor-based ML models, and (iii) a combined ML-augmented RK approach (ML4RK). Leave-one-element-out tests highlight complementary strengths. RK models (class (i)) remain the most data-efficient when binary information is available, while descriptor-based ML models (class (ii)) enable genuine zero-shot extrapolation to elements absent from the training set. By embedding elemental descriptors into the RK framework, the hybrid approach unifies these regimes and enables some level of prediction of interaction parameters for otherwise unknown or data-scarce binaries (class (iii)). This work demonstrates a physically grounded and data-efficient route to extend CALPHAD models, combining the transferability of ML with the physical grounding, interpretability, data efficiency, and robustness of thermodynamic formalisms.
}


\maketitle

\section{Introduction}\label{sec1}

Accurate phase diagram calculations are central to understanding thermodynamic behavior and are essential for alloy design, process optimization, and predicting phase stability across conditions. The CALPHAD (CALculation of PHAse Diagrams) method has long provided a rigorous and extensible framework for modeling such equilibria~\cite{liu2023thermodynamics}. Since its development in the 1970s, CALPHAD has enabled analytical descriptions of the Gibbs free energy of multicomponent phases as functions of temperature, pressure, and composition~\cite{kattner1997thermodynamic}. These models are parameterized in a hierarchical manner, beginning with unary and binary systems and extending to ternary and higher-order systems. With support from mature software platforms such as Thermo-Calc~\cite{andersson2002thermo}, Pandat~\cite{chen2002pandat}, and OpenCalphad~\cite{sundman2015opencalphad}, CALPHAD has facilitated the development of highly effective thermodynamic databases for metals and ceramics. These databases can model very complex multicomponent systems, including both common single-element dominated alloys (e.g., Fe-, Al-, and Ni-based alloys) and increasingly so-called compositionally complex (also called high-entropy) materials (e.g., high-entropy alloys, oxides, carbides, and borides) ~\cite{akrami2021high,zeng2021revealing,sulley2024accelerating,bansal2023accelerated,kaufmann2020searching,jiang2023current,zhang2024determination,liu2022design,pei2020machine,vazquez2023deep,liu2024comparative,zhu2024accelerating,qi2019high,zhang2022calphad}. These databases are grounded in standardized elemental reference data from SGTE~\cite{dinsdale1991sgte} and now form a core component of modern computational materials design workflows.

Despite this success, CALPHAD faces several inherent limitations. Constructing reliable thermodynamic descriptions for higher-order systems remains labor-intensive, often requiring extensive experimental data and expert optimization. As a result, most established databases focus on fitting to unary, binary, and a limited number of ternary systems. In addition, CALPHAD models are restricted by their analytical functional forms and can only predict phases for which thermodynamic assessments already exist, limiting their applicability to new or poorly characterized chemistries. These challenges are particularly pronounced for emerging materials such as multi-principal element metal alloys, oxides, carbides, etc. (also called High-Entropy Alloys (HEAs)), where data scarcity is common but predictive capability is crucial.

Recent advances in machine learning (ML) provide opportunities to mitigate these limitations and broaden the reach of CALPHAD modeling~\cite{shen2025synergy}. ML has been used to accelerate data acquisition, optimize model parameters, quantify uncertainty, and rapidly estimate thermodynamic quantities, such as Gibbs free energies, chemical potentials, and phase boundaries. Physics-informed ML approaches that embed thermodynamic constraints directly into the model have further improved robustness and consistency. These developments extend CALPHAD’s predictive capability into data-scarce regions and open pathways for inverse design, active learning, and autonomous exploration of composition–temperature spaces. Consequently, ML is evolving from an auxiliary tool into a central component of next-generation thermodynamic modeling.

In this work, we focused on the question: \textit{How Can Machine Learning Accelerate CALPHAD Free Energy Modeling?} By this, we mean that we considered the actual process of fitting free energy models. It is tempting to simply replace the traditional CALPHAD functional forms with ML models, e.g., as has been done in Refs.~\cite{shen2025synergy,hong2023deep}. However, we show that, at least for the data considered here, this is not the optimal approach, as the physically motivated Redlich–Kister (RK) polynomials provide excellent data-efficient fits. The ability of a model's fitting function's form to provide inherent constraints is sometimes called the "structural bias" of the model. Therefore, we can conclude that, except for the most realistic data sets, the structural bias of RK polynomials is very effective for free energy modeling, consistent with decades of successful utilization of RK polynomials in CALPHAD models. However, there are still opportunities for using ML even with direct RK polynomial fits. One simple idea that we demonstrated here is regularizing the fitting of the RK coefficients using the Least Absolute Shrinkage and Selection Operator (LASSO), as this produces high-quality fits while choosing just the required polynomial terms from a large set of options. A deeper idea explored here is to utilize the ability of ML functions to learn on features besides composition to support greater extrapolation than traditional RK approaches.

Instead of directly replacing the RK polynomials with ML functions, we developed here a machine-learning–augmented CALPHAD strategy that learns RK interaction coefficients directly from physically informed elemental descriptors. This approach allows us to retain the structural bias and thermodynamic consistency of the RK formalism, while leveraging the generalization capability of ML to estimate interaction parameters for element pairs that are not present in the training data. A related idea has previously been explored for liquid alloys~\cite{hong2023deep}; however, their neural-network model replaces the polynomial form entirely, which reduces interpretability and makes extrapolation less transparent within the CALPHAD framework. In contrast, we adopt a more physics-constrained and interpretable formulation for multi-elemental solid solution phases: ML is used only to infer the composition-independent RK coefficients from elemental descriptors, while the classical RK polynomial structure is preserved to build composition-dependent Gibbs energies. This keeps the CALPHAD workflow intact, provides clear physical meaning to the learned parameters, and enables principled extensibility without abandoning the underlying thermodynamic model. It also has the advantages that (i) the final models can be expressed in standard CALPHAD form, stored in Thermodynamic Database (TDB) files, and used in standard codes (e.g., pycalphad, Thermo-Calc, Pandat, etc.) and (ii) the approach can be trivially generalized to ML of parameters in any other CALPHAD function forms, e.g., parameters in the Einstein model for vibrations or the Modified Quasichemical Model for liquids.

Our assessment is done using the composition-dependent formation energies of 14-element FCC alloys with random chemical order (this ordering approximates a solid solution), computed with a universal ML interatomic potential. Details on the data (e.g., size, compositions, and calculation methods) are in the Methods Section, and the data availability is described in the "Data and code availability" section. By using computed data for a wide range of elements, we can explore the numerical aspects of different CALPHAD and ML approaches in far more detail than is possible with experimental data alone. We systematically compare three approaches to modeling these energies: (i) RK and ML models trained solely on composition features, (ii) ML models trained on elemental descriptors, and (iii) a hybrid framework that predicts RK coefficients using ML. Approach (i) is the natural one to consider first, and it assesses the effectiveness of simply replacing RK with ML while keeping the descriptors fixed as composition. Approach (ii) explores changing the descriptors from composition to elemental descriptors, which is possible with ML but not RK methods, and is explored as it might improve the ML and could allow for more extrapolation to new elements. We find that with composition features, RK models are highly data-efficient and converge rapidly, generally outperforming ML methods, particularly when binary information is available. However, we find that descriptor-based ML models enable zero-shot extrapolation to unseen elements, but they cannot be used in traditional CALPHAD RK fits.  Our approach (iii), which we call ML4RK, integrates the best aspects of both approaches (i) and (ii), using the RK functional form and the elemental featurization together. We show that ML4RK allows for reliable fitting as well as some level of extrapolation to compositions containing elements with little or no data. Together, these results demonstrate that learning physically informed RK coefficients with ML provides a data-efficient and extrapolatable approach to model free energies. We stress that ML4RK uses traditional CALPHAD fitting as its first step, and therefore shares all its strengths. ML4RK then provides additional capabilities through ML, giving it clear advantages over traditional CALPHAD approaches alone.

\section{Results}\label{sec2}

\subsection{Reference and training datasets}\label{sec2.05}
In this and the following sections, we will evaluate performance using three reference datasets. They are

(i) \textit{all-14-Multi}: 1000 entries for 14-component alloys;

(ii) \textit{all-$C_{14}^2$-Binary}: 3185 entries for binary systems, where each of the 91 binary combinations among 14 elements contains 35 data points ($35\times91=3185$);

(iii) \textit{all-13-Multi}: 1000 entries for 13-component alloys.

Set (i) and (ii) allow assessment of multicomponent and binary trends, respectively, while set (iii) allows for assessment of elements not included in the training data. See Method Section for more details on these datasets.

We first evaluated RK and ML models using composition fractions as the sole feature set (approach (i) in the Method Section). To probe data dependence, we constructed training subsets of increasing size from the reference datasets and considered three training schemes: 

(i) \textit{Training-1 (14-component alloys)}, where 50–800 samples were drawn from the \textit{all-14-Multi} dataset, with the 200 remaining compositions not included in the 800-sample set forming the \textit{20\%-14-Multi} test set; 

(ii) \textit{Training-2 (binary alloys)}, where subsets containing 2–28 samples per binary system were selected from the \textit{all-$C_{14}^2$-Binary} dataset and compared against the \textit{20\%-$C_{14}^2$-Binary} test (note that this represents 7 values, or 20\%, of the data for each binary alloy chosen at random, thus ensuring that all binary alloys are equally represented in the test data); 

(iii) \textit{Training-3 (combined datasets)}, where binary and multicomponent subsets of similar size were paired to create systematically growing mixed training sets. 

\subsection{RK polynomial and ML model fitting based on composition features}\label{sec2-1}

\begin{figure}[ht!]
\centering
\includegraphics[width=\textwidth]{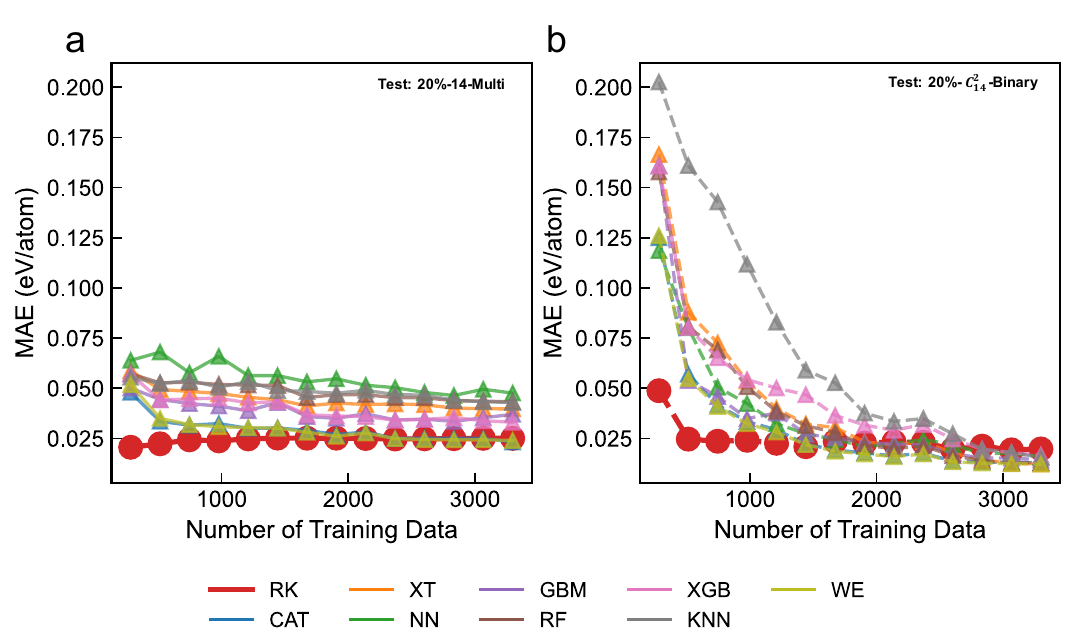}
\caption{
\textbf{Comparison of RK and ML model performance based on composition features.}
Prediction accuracy (MAE, eV/atom) versus training-set size for the Redlich–Kister (RK, red circles) and multiple machine-learning (ML) models under two train–test configurations.
Training on the combined dataset [80\%-(Multi+Binary)]; tests on (a) 20\%-14-Multi and (b) 20\%-Binary.
Across panels, RK shows strong data efficiency and robust transfer across distribution shifts, whereas ML curves tend to plateau earlier. ML model abbreviations: XT = ExtraTrees; GBM = LightGBM; NN = NeuralNetTorch; XGB = XGBoost; RF = RandomForest; KNN = KNeighbors; CAT = CatBoost; WE = WeightedEnsemble.
}

\label{fig1}
\end{figure}

For each training set described in Sec~\ref{sec2.05}, we fitted a Lasso regression model using RK-like features and trained AutoGluon ML models~\cite{erickson2020autogluon} on the same composition inputs (see Method Section for more details on the ML modeling). Full learning curves and hyperparameter details for Training-1 and Training-2 are provided in the Supplementary Information (Fig.~S1).

In the main text, we focus on the combined datasets from \textit{Training-3} (Fig.~\ref{fig1}), since this setting represents a general case where binary and multicomponent information jointly constrain interaction parameters. When both data types are available, RK consistently achieves the lowest MAEs on both \textit{20\%-$C_{14}^2$-Binary} and \textit{20\%-14-Multi} tests as the training size increases. ML models improve as more data are added, but their errors remain above those of RK across the full range of \(N\). This reflects RK’s strong pairwise structural bias: once sufficient binary information is present, the functional form efficiently recovers the dominant interaction trends and generalizes well to higher-order compositions. In contrast, ML models seem to require large amounts of training data to learn the dominant contribution and be able to effectively predict the test data. ML models do outperform RK when learning from only multicomponent data \textit{Training-1} and predicting binary alloys (see Fig.~S1(b)), but this is not a case we expect to see often, given the typical way complex alloy thermodynamic models are built from suballoys.

Taken together, these results show that composition-based RK remains highly competitive and often superior when the training data include informative binary interactions. Binary subsets in particular allow RK to fully leverage its pairwise structure, yielding robust and data-efficient predictions even in multicomponent test environments. However, composition-only ML models lack a chemically meaningful metric between elements, since they have no natural correlation between some measure of their distance and their properties. They are likely to do better with some continuous embedding that relates meaningfully to elemental behavior. Furthermore, composition-based RK and ML models cannot extrapolate to new elements since they have no way to relate a new element to those they have seen. We can overcome these limitations by the introduction of physically informed elemental descriptors, described in the next section.

\subsection{Can physically informed elemental descriptors improve ML beyond the RK polynomial?}

\begin{figure}[ht!]
\centering
\includegraphics[width=\textwidth]{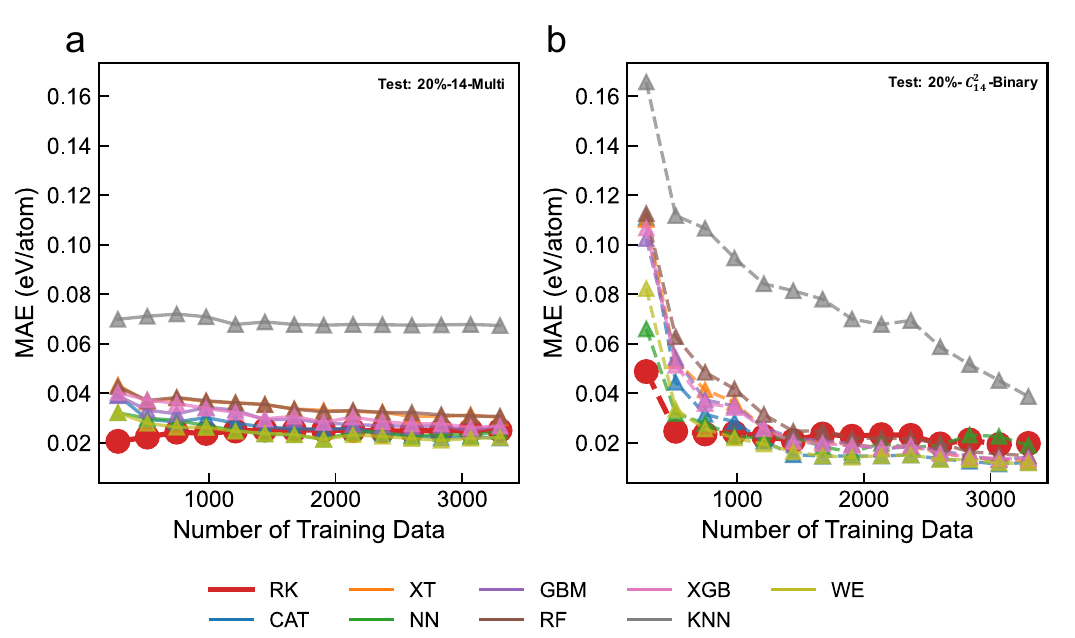}
\caption{\textbf{Comparison of RK and ML model performance based on elemental features.}
Prediction accuracy (MAE, eV/atom) versus training-set size for the Redlich–Kister (RK, red circles) and multiple machine-learning (ML) models under two train–test configurations.
Training on the combined dataset [80\%-(Multi+Binary)]; tests on (a) 20\%-Multi and (b) 20\%-Binary.
Across panels, RK shows strong data efficiency and robust transfer across distribution shifts, whereas ML curves tend to plateau earlier. ML model abbreviations: XT = ExtraTrees; GBM = LightGBM; NN = NeuralNetTorch; XGB = XGBoost; RF = RandomForest; KNN = KNeighbors; CAT = CatBoost; WE = WeightedEnsemble.
}
\label{fig2}
\end{figure}

To assess whether physically informed elemental descriptors enhance ML performance, we replaced raw composition fractions with Magpie-based \texttt{ElementProperty} features~\cite{ward2016general,ward2018matminer}. Fig.~\ref{fig2} presents the results for the mixed-data setting (\textit{Training-3}), which best reflects realistic thermodynamic-model development. Full results for \textit{Training-1} and \textit{Training-2} are provided in the SI and show analogous trends (Fig.~S2). Note that in all these figures the RK fits still use element composition as their features since they are not formulated to use elemental properties. 

With descriptor-based features and mixed training data, ML models exhibit markedly improved accuracy and generalization. On the \textit{20\%-14-Multi} test, ML curves closely track RK, and ensemble models slightly outperform RK at large $N$. On the \textit{20\%-$C_{14}^2$-Binary} test, ML performance nearly overlaps with RK, eliminating the clear ML–RK gap observed when only composition features were used.

Overall, these results show that physically informed elemental descriptors greatly enhance the representational capacity of ML models. While RK remains highly efficient for binary-focused learning due to its pairwise functional form, descriptor-based ML models now approach or match RK in most regimes and significantly outperform composition-based ML models when evaluated outside the distribution on which they were trained. The \texttt{ElementProperty} featurizer, therefore, closes much of the performance gap between ML and RK, highlighting the value of incorporating physically meaningful attributes when predicting thermodynamic properties.

\subsection{Extending to a new alloying element: zero-shot extrapolation and few-shot learning}

\begin{figure}[H]
\centering
\includegraphics[width=0.85\textwidth]{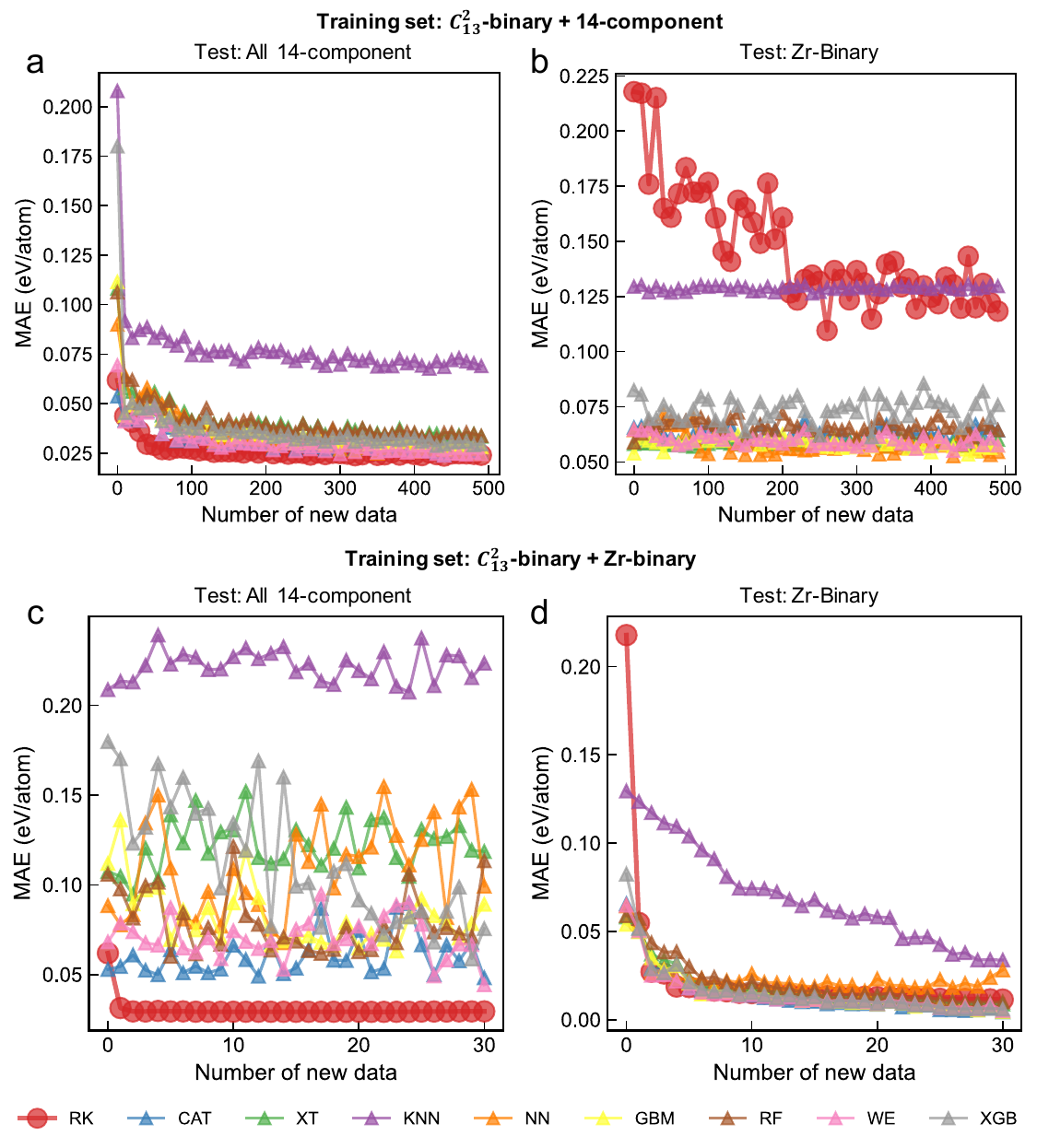}
\caption{\textbf{Zero-shot and few-shot learning when adding Zr as a new element.}
Rows correspond to four training configurations that exclude Zr initially and then add Zr-containing data: 
training data from all $C_{13}^{2}$ binaries plus added 14-component points, then test on (a) Test: All 14-component and (b) Test: Zr-Binary; 
training data from all $\binom{13}{2}$ binaries plus added Zr–binary points, then test on (c) Test: All 14-component and (d) Test: Zr-Binary. Red circles denote RK; other curves are ML models. ML model abbreviations: XT = ExtraTrees; GBM = LightGBM; NN = NeuralNetTorch; XGB = XGBoost; RF = RandomForest; KNN = KNeighbors; CAT = CatBoost; WE = WeightedEnsemble.}
\label{fig3}
\end{figure}

To understand how CALPHAD and ML models behave when a \emph{new} alloying element is introduced, we adopt a leave-one-element-out (LOOE) protocol in which all binaries and multicomponent compositions containing the held-out element are removed from training. The predictive task is then to estimate formation energies for compositions containing the new element, either without any prior data (\emph{zero-shot}) or after adding a controlled number of samples containing the element (\emph{few-shot}). This setting mirrors the practical challenge of extending a thermodynamic database when a new solute has no or limited experimental characterization.

Fig.~\ref{fig3} presents two of the training cases, both of which begin from a fully constrained non-Zr binary space. In the first configuration, the model is trained on all $C_{13}^{2}$ binaries and then progressively augmented with Zr-containing 14-component data. In the second configuration, the same $C_{13}^{2}$ binary backbone is instead augmented with targeted Zr–binary measurements. Results from the analogous 13-component–based configurations are included in the Supplementary Information (Fig.~S3), where they exhibit similar qualitative behavior but involve weaker initial constraints due to the absence of binary anchoring.

Across both training sets in Fig.~\ref{fig3}, the same zero-shot pattern emerges. Without any Zr-containing data ($N=0$), descriptor-based ML models significantly outperform RK on all Zr–binary predictions. This is expected: the RK framework cannot infer missing $L_{\mathrm{Zr-X}}$ coefficients and defaults to assigning no interaction term, whereas ML models place Zr within a chemically meaningful descriptor space and infer approximate behavior by analogy with other elements. Even though the non-Zr binary network strongly constrains interactions among the remaining 13 elements, any composition involving Zr inevitably depends on unconstrained parameters in the RK approach, leaving RK systematically biased until Zr-specific data are supplied.

The learning trajectories behave differently depending on the type of Zr-containing samples that are introduced. When the added data consist of Zr–binary measurements (as shown in Fig.~\ref{fig3}c and d), RK responds immediately and reduces rapidly: even a handful of Zr–binary points reduces RK’s Zr–binary MAE by more than an order of magnitude, after which RK becomes the most accurate model on the binary test. These few binary anchors also propagate efficiently into multicomponent predictions, rapidly driving RK toward ML-level accuracy on 14-component alloys. ML models also benefit from Zr–binary additions, but their improvement is more gradual, reflecting their weaker structural bias toward pairwise mixing physics.

In contrast, when the added data consist only of Zr-containing 14-component compositions (as shown in Fig.~\ref{fig3}a and b), the signal is weaker. Multicomponent samples partially constrain Zr interactions because Zr appears in diverse local environments, but each sample provides only diffuse information that averages over many competing pairs. As a result, both RK and ML improve with additional 14-component points, but more slowly than with binary information. Interestingly, errors for the best ML models and RK largely converge for the full 14-component test set (Fig.~\ref{fig3}a), but the Zr-Binary test set (Fig.~\ref{fig3}b) shows that the Zr-binary alloys are never well represented by the RK model.

Taken together, these results reveal a clear division of strengths. Descriptor-based ML offers the most reliable zero-shot predictions for completely unseen elements, leveraging periodic trends to infer interaction behavior in the absence of direct data. The RK framework, by contrast, is the most data-efficient learner once targeted binary information is available, rapidly achieving the best performance across both binary and multicomponent systems. These complementary behaviors suggest a practical pathway for database development: use elemental features to get chemically informed initial estimates for new elements, but use RK to deliver rapid, thermodynamically consistent refinement as soon as key binary measurements are acquired. To achieve this, we need an approach that can integrate the composition dependence of RK polynomials with elemental-property descriptors. We describe one such approach below.

\subsection{An ML-augmented Redlich--Kister framework that learns interaction coefficients from elemental descriptors (ML4RK)}

A major limitation of conventional CALPHAD modeling is that Redlich--Kister (RK) interaction parameters cannot be estimated for binary systems for which no thermodynamic data are available. This restriction leaves large regions of chemical space inaccessible and prevents the construction of thermodynamic descriptions for chemically novel or experimentally challenging binaries. To address this limitation while retaining the structural advantages and interpretability of the RK formalism, we propose a machine-learning-augmented Redlich--Kister framework, hereafter referred to as ML4RK. The central idea is to learn RK interaction coefficients from physically informed, composition-independent elemental descriptors. This enables quantitative extrapolation of RK coefficients to completely unseen binaries while preserving the standard composition dependence of the RK expansion.

We demonstrate the approach in two steps. First, all available binary formation-energy data are fitted using a global RK regression, in which the formation energy of each binary sample is represented as
\[
\Delta H_{ij}(x_i,x_j)
=
x_i x_j \sum_{v=0}^{V_{\max}}
L_{ij}^{(v)} (x_i-x_j)^v .
\]
This regression yields a reference set of RK coefficients, $\{L_{ij}^{(v)}\}$, for all binaries covered by the dataset. These RK coefficients serve as supervised targets for the subsequent machine-learning stage.

Second, we construct a composition-independent descriptor vector for each binary pair and learn the RK coefficients for each order $v$ independently. For a given binary pair $i$--$j$, the descriptor map is evaluated at the equiatomic composition, $x_i=x_j=0.5$. Using Matminer’s Magpie framework~\cite{ward2016general,ward2018matminer}, each binary pair is represented by Magpie elemental-property statistics computed for the equiatomic pseudo-composition. The resulting descriptor vector includes statistics such as the mean, range, maximum, minimum, mode, mean absolute deviation, and average deviation of elemental properties across the two end members. These elemental properties include quantities related to atomic size, atomic mass, periodic-table position, electronegativity, valence-electron configuration, melting temperature, and other chemically meaningful descriptors. Because the descriptor is evaluated at the equiatomic composition, it depends only on the identity of the two elements and not on the particular alloy composition being predicted. We denote this binary-pair descriptor vector as $\boldsymbol{z}_{ij}$.

For each RK order $v$, a separate regression model $\mathcal{M}_v$ is trained to learn the mapping from the binary-pair descriptor vector to the corresponding RK coefficient:
\[
\widehat{L}_{ij}^{(v)}
=
\mathcal{M}_v\!\left(\boldsymbol{z}_{ij}\right).
\]
Tree-based ensemble models, including LightGBM, XGBoost, and CatBoost, are well suited to this task because they can capture nonlinear descriptor--coefficient relationships while remaining computationally efficient and interpretable through feature-importance and SHAP analyses. In the present study, we adopt CatBoost as the representative regressor in the ML4RK framework because it delivers consistently strong predictive performance across the benchmark tests performed here.

Once trained, these models can produce RK coefficients for any binary pair for which elemental descriptors are available:
\[
\widehat{L}_{ij}^{(v)}
=
\mathcal{M}_v\!\left(\boldsymbol{z}_{ij}\right).
\]
The predicted coefficients are then inserted into the RK expansion to obtain the formation energy at any composition:
\[
\widehat{\Delta H}_{ij}(x_i,x_j)
=
x_i x_j \sum_{v=0}^{V_{\max}}
\widehat{L}_{ij}^{(v)}(x_i-x_j)^v .
\]
In this formulation, the ML model does not directly learn a composition-dependent formation-energy surface. Instead, it learns composition-independent RK interaction coefficients from elemental descriptors, while the RK polynomial provides the full composition dependence. In this way, ML4RK preserves the structural transparency of the CALPHAD formalism while learning how effective interaction strengths vary across chemical space. Physically, the model infers the magnitude and sign of binary mixing energetics---that is, whether a system tends toward ordering or phase separation---from systematic trends in elemental size mismatch, cohesion-related properties, valence-electron characteristics, electronegativity, and other chemically meaningful descriptors.

\begin{figure}[H]
\centering
\includegraphics[width=0.8\textwidth]{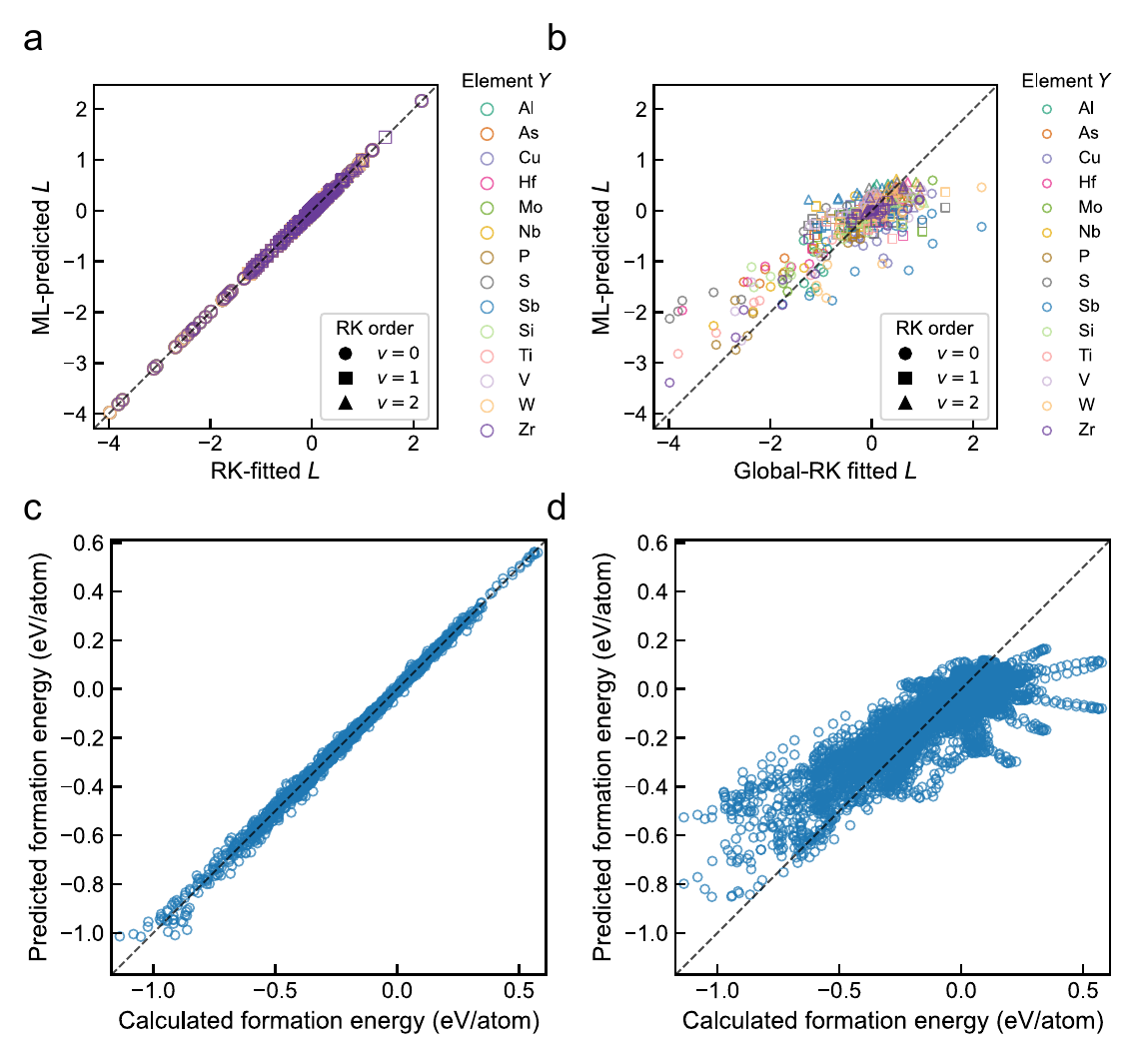}
\caption{
\textbf{Performance of the machine-learning-augmented Redlich--Kister (ML4RK) framework.}
(a) Comparison between RK-fitted interaction coefficients $L_{ij}^{(v)}$ and ML-predicted coefficients for binaries included in the coefficient-learning stage. The ML models use composition-independent binary-pair descriptors evaluated at the equiatomic composition, and each RK order $v$ is learned by a separate regressor. Points correspond to element pairs and RK orders, with clustering along the diagonal indicating successful learning of the elemental descriptor--coefficient mapping. (b) Leave-one-element-out validation of the predicted RK coefficients. For each excluded element $Y$, all $Y$--$X$ binaries are removed from training, and the corresponding coefficients $L_{YX}^{(v)}$ are predicted from the composition-independent descriptors of the omitted binary pairs.(c) Parity plot for the global RK regression, showing that the RK polynomial expansion accurately reproduces the available binary formation energies. (d) Zero-shot formation-energy predictions for all omitted $Y$--$X$ binaries in the leave-one-element-out tests. Although these binary systems are completely absent from the training data, the ML-predicted RK coefficients, when inserted into the RK expansion, yield formation energies that follow the reference values. RMSE = 0.1250 eV/atom. Additional prediction metrics, including mean absolute error (MAE), root mean square error (RMSE), and the coefficient of determination ($R^2$), are provided in the Supporting Information.
}
\label{fig:lgbm_overall}
\end{figure}

Fig.~\ref{fig:lgbm_overall} summarizes the performance of the ML4RK framework on our dataset. The ML-predicted coefficients for the training binaries (Fig.~\ref{fig:lgbm_overall}a) cluster closely around the RK-fitted values for all RK orders, demonstrating that the mapping from composition-independent binary-pair descriptors to RK interaction coefficients, $\widehat{L}_{ij}^{(v)}=\mathcal{M}_v(\boldsymbol{z}_{ij})$, is learned effectively within the known chemical space. The global RK fit (Fig.~\ref{fig:lgbm_overall}c) reproduces the available binary formation-energy data with an RMSE on the order of $10^{-2}$~eV/atom, confirming that the RK expansion provides an accurate and compact representation of the underlying thermodynamic data. The more stringent test is the leave-one-element-out validation shown in Fig.~\ref{fig:lgbm_overall}b, where all binaries containing a selected element $Y$ are removed from training, and the corresponding $Y$--$X$ interaction coefficients are predicted only from their binary-pair descriptors. The best correlation is observed for the $v=0$ coefficients, which provide the dominant contribution to the mixing energetics. Higher-order coefficients show larger relative scatter, reflecting their smaller magnitudes and stronger sensitivity to details of the fitted composition dependence. However, deviations in these higher-order terms have a reduced influence on the final formation energies because their contributions are weighted by the RK basis factors $(x_i-x_j)^v$. Fig.~\ref{fig:lgbm_overall}d shows the resulting zero-shot formation-energy predictions for the same omitted $Y$--$X$ binaries. In this case, the ML-predicted RK coefficients are inserted into the RK expansion and evaluated across the corresponding binary compositions. The predicted formation energies remain strongly correlated with the reference values, even though these binaries are entirely absent from the coefficient-learning stage. This result demonstrates that ML4RK can extrapolate thermodynamic trends to unseen binary systems while retaining the physically structured composition dependence of the RK formalism.

\begin{figure}[H]
\centering
\includegraphics[width=\textwidth]{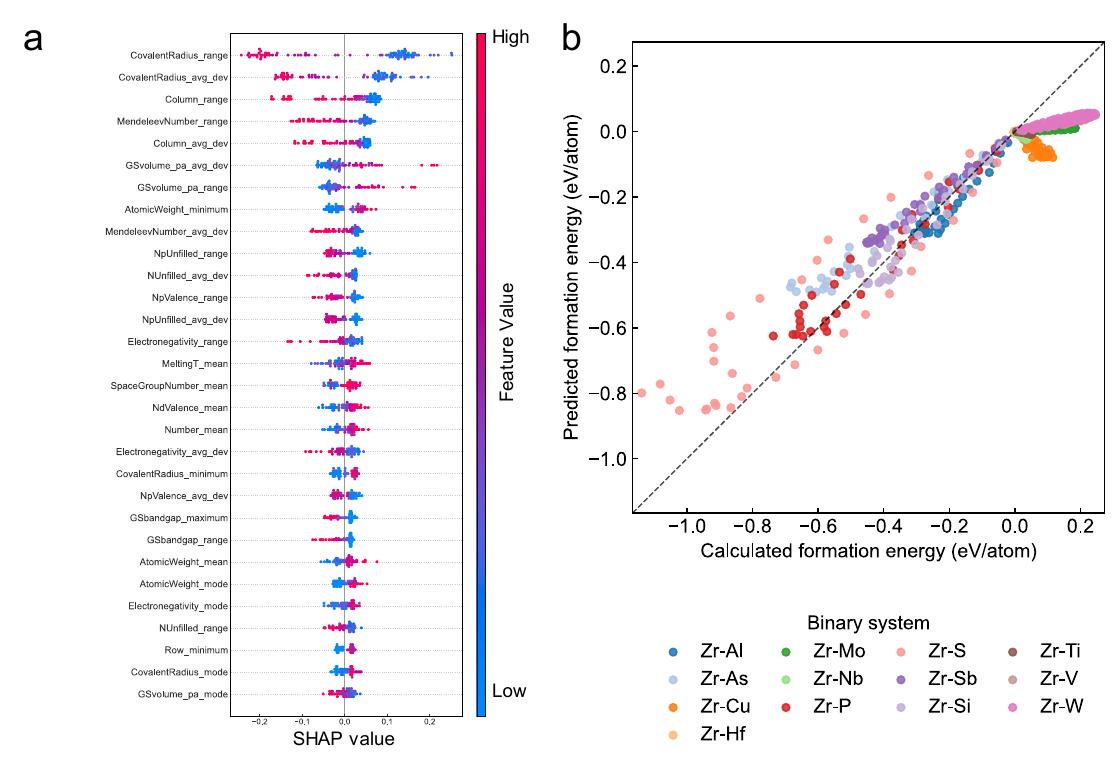}
\caption{
\textbf{Case study of Zr–X binaries: zero-shot extrapolation and descriptor-based interpretability.}
(a) SHAP analysis identifying the elemental descriptors most influential for predicting $L_{Zr-X}^{(v)}$, with size-related features (covalent radius, atomic weight, atomic volume) and chemical descriptors.
(b) Zero-shot formation energies for all Zr–X binaries obtained by inserting ML-predicted interaction parameters into the RK expansion. The predictions align well with reference values. More information on prediction quality, evaluated using three standard regression metrics—mean absolute error (MAE), root mean square error (RMSE), and the coefficient of determination ($R^2$)—is provided in the Supporting Information. The full list of SHAP feature labels and their physical meanings is provided in Supplementary Table~S5.
}
\label{fig:ZrX}
\end{figure}

To examine the physical meaning encoded in the learned interaction parameters, we perform SHAP analysis using models trained on the full binary dataset, complemented by a representative case study of Zr--X binaries (Fig.~\ref{fig:ZrX}). The SHAP beeswarm plot (Fig.~\ref{fig:ZrX}a) summarizes feature importance across the training samples and provides a chemically interpretable view of the factors governing the RK interaction coefficients. Several classes of descriptors emerge as consistently influential. Size- and packing-related quantities, including the range and average deviation of covalent radius, the range and average deviation of ground-state atomic volume, and the minimum atomic weight, dominate the importance ranking. These descriptors directly reflect atomic-size mismatch, packing contrast, and mass-related differences between the two elements, all of which are closely connected to local strain and the magnitude of mixing enthalpies. Chemical-scale descriptors, including the range of periodic table columns, the range of Mendeleev number, and the range of electronegativity, also contribute strongly. Their importance indicates that the learned interaction parameters are sensitive to elemental separation across the periodic table and to differences in bonding tendency. In addition, valence-related descriptors, such as the range and average deviation of \(p\)-valence electrons and the number of unfilled electronic states, capture variations in electronic configuration and available bonding states. Together, these feature groups correspond closely to the classical physical controls of alloy thermodynamics: geometric strain, chemical contrast, electronic compatibility, and cohesive strength.

Overall, the feature trends are chemically reasonable. For example, a large difference in periodic-table column or Mendeleev number generally promotes more negative \(L_{ij}\) values and formation energies, consistent with stronger heteroatomic bonding driven by chemical contrast and charge transfer. One notable trend is associated with the range of covalent radius, which has a strong contribution for systems with a large size mismatch. At first sight, this may appear to contradict the simple Hume--Rothery size-mismatch argument, where large atomic-size differences are expected to increase the strain penalty and destabilize mixing. However, in the present dataset, the largest size-mismatch systems are often not simple metallic alloys. For example, Zr--P combines substantial size mismatch with strong chemical and electronic contrast, leading to strong heteroatomic bonding and highly negative formation energies. Therefore, the SHAP trend associated with covalent-radius mismatch should not be interpreted as a pure elastic-size effect. Instead, it reflects the coupled influence of size mismatch, charge transfer, and bond strength in the available chemical space. The Zr--X parity plot in Fig.~\ref{fig:ZrX}b is consistent with this interpretation. The model captures the broad energetic ordering across Zr-containing binaries, including strongly negative systems such as Zr--P and weaker interactions among transition-metal pairs. The remaining deviations mainly occur in systems where several effects compete, such as size mismatch, chemical separation, and electronic-structure differences. These results indicate that the ML4RK model learns physically meaningful descriptor--coefficient relationships rather than arbitrary statistical correlations.

These results show that the ML4RK framework can perform extrapolation to unseen binaries. The combination of the RK formalism with descriptor-based ML thus provides a thermodynamically consistent and data-efficient pathway for extending CALPHAD models into regions of chemical space where direct assessments are unavailable or severely limited. The accuracy of these predictions is still limited, and further work is needed. See the Discussion Section for further discussion.

\section{Discussion}\label{sec12}

This work uses machine learning and physics-based elemental descriptions to develop an approach to extend CALPHAD assessments into regions of chemical space where direct experimental or assessed data are missing or scarce By disentangling the strengths and limitations of composition-based RK models, descriptor-based ML models, and their hybrid integration, we demonstrate that these approaches are not competing alternatives but complementary tools whose capabilities map naturally onto different stages of database construction.

Our results establish three overarching conclusions. First, CALPHAD’s long-standing RK formalism remains exceptionally data-efficient when binary information is available. In few-shot regimes, especially when targeted binary measurements for a new element are provided, RK consistently yields the most accurate and stable predictions. Second, descriptor-based ML models enable “zero-shot’’ extrapolation, inferring mixing energetics for completely unseen elements by leveraging physically meaningful trends encoded in elemental features. This capability is inherently absent in traditional RK parameterizations, which cannot infer $L_{ij}^{(v)}$ without explicit data. Third, the proposed ML4RK framework unifies these strengths: ML models learn how interaction parameters vary across chemical space, while RK preserves thermodynamic consistency and interpretability. As a result, the hybrid approach can populate previously inaccessible regions of binary interaction space and deliver CALPHAD-ready parameters before experimental data are available.

To complement the above model-generated 14-element FCC data, we also performed leave-one-out and ML4RK tests using real thermodynamic data extracted from a CALPHAD database at 1000~K (as discussed in the Supplementary Information). These tests broadly reproduce the qualitative behaviors observed in the synthetic dataset: ML provides some zero-shot performance not available to RK fits, and RK rapidly becomes accurate once targeted binary data are included. However, the quantitative performance differs from the model-generated system. Several elements lack a stable FCC phase at 1000~K, resulting in incomplete or sparse data for specific binaries. This and other sources of noise mean that the underlying RK parameters exhibit variable quality. As a consequence, the ML4RK fitting accuracy on real data is lower than in the simulated dataset. These experiments reinforce an important point: real-world thermodynamic assessments involve domain-specific biases (e.g., phase stability, solution model choices, and data availability) that are absent from idealized benchmark systems.

It is important to emphasize that the goal of this work is not to produce a final, highly optimized surrogate model, but rather to establish and validate a generalizable concept: that RK interaction parameters can be learned, extrapolated, and physically interpreted directly from elemental descriptors. As such, we intentionally adopt standard ML approaches rather than the most advanced architectures or task-specific optimization. We expect that combining the ML4RK strategy with stronger modern regressors—including graph neural networks, attention-based models, or foundation materials models—together with active-learning-based sampling of high-value binary and multicomponent points, will significantly improve quantitative accuracy and robustness of predicted $L_{ij}^{(v)}$. In particular, iterative acquisition of targeted binary measurements could close remaining uncertainties in $L_{ij}^{(v)}$ and enable the automated construction of thermodynamic databases that generalize reliably across the periodic table.

It should be emphasized that the present dataset is designed as a controlled benchmark for evaluating thermodynamic-model construction strategies, rather than as a final CALPHAD assessment of the selected chemical systems. The target quantities are 0 K formation energies computed from a universal machine-learning interatomic potential for ideal FCC substitutional configurations with volume relaxation only. These energies therefore represent the energetic component of a CALPHAD-like description and do not include vibrational, magnetic, electronic, or configurational entropy contributions. Consequently, the present results should be interpreted as a demonstration of how elemental descriptors can be used to infer Redlich–Kister interaction parameters and extend composition-dependent energy models across chemical space. Extension to full Gibbs-energy assessments will require temperature-dependent terms, reference-state consistency, phase-stability constraints, and validation against experimental or assessed thermodynamic data.

Finally, this study reframes how CALPHAD and ML can be integrated: ML provides a framework that can utilize chemically meaningful features that enable extrapolation to new elements, while the RK formalism supplies a physically grounded and efficient structure for thermodynamic function modeling and interpolation. This division of labor offers a new pathway toward data-efficient expansion of thermodynamic databases, opening possibilities for accelerated alloy design, autonomous CALPHAD assessments, and large-scale exploration of multicomponent phase stability.

\section{Methods}\label{sec11}
\subsection{Data generation}
To construct a diverse set of multi-component FCC alloy chemical configurations, we employed the Atomic Simulation Environment (ASE) framework~\cite{larsen2017atomic}. Starting from an ideal face-centered cubic (FCC) unit cell of elemental Fe, a $3 \times 3 \times 3$ supercell was generated, yielding 108 atomic sites. The alloying elements were selected from a set of 14 species (Mo, V, W, Si, Cu, Ti, Al, S, P, Zr, Nb, As, Sb, and Hf), chosen to represent the compositional diversity commonly explored in high-entropy alloys (HEAs) and complex concentrated alloys (CCAs). This set spans transition metals, refractory elements, and light non-metals, thereby encompassing a wide range of atomic sizes, valence electron counts, and bonding characteristics known to stabilize disordered solid solutions. Strongly magnetic elements such as Fe, Co, and Ni were deliberately excluded, as universal machine-learning interatomic potentials typically provide more reliable descriptions of non-magnetic interactions, while magnetism can introduce systematic errors.

For 13- and 14-component alloys, compositions were sampled from a Dirichlet distribution with uniform concentration priors~\cite{ng2011dirichlet}, ensuring random yet normalized mole fractions across all elements for each structure. The resulting atom counts were rounded to integers with corrections applied to preserve the total atom number, and the final element assignments were randomly permuted before being mapped onto the atomic sites of the supercell. For binary alloys, compositions were generated using a chunk-based enumeration scheme rather than random sampling. Each 108-atom FCC supercell was divided conceptually into 36 composition chunks, with each chunk corresponding to three atoms. The number of chunks assigned to the first element was varied from 1 to 35, and the second element occupied the remaining chunks. This construction produced binary compositions spanning the full range in uniform increments of $3/108$, excluding the two pure end-member cases. For each resulting composition, the atomic identities were randomly shuffled before assignment to the FCC lattice sites, thereby generating disordered solid-solution structures at each discrete composition point.

To compute the formation energy of each high-entropy alloy structure, we relaxed volume but did not relax the cell parameter lengths independently or their angles, and we did not relax local atomic positions. This was necessary to avoid many instabilities on the FCC lattice associated with the many chemistries explored. While this means that the energies are not very accurate values for true FCC alloys of these elements, they still represent a collection of highly physical energies suitable for assessing different free energy modeling approaches. 

We employed a two-step volume optimization procedure based on energy–volume (E–V) curves, using a pre-trained machine learning interatomic potential (MLIP), MatterSim-v1.0.0-5M.pth~\cite{yang2024mattersim}, without additional fine-tuning. Each structure was constructed using the ASE and assigned the MatterSim potential through its Python calculator interface. An initial E–V curve was generated by isotropically scaling the simulation cell from 80\% to 120\% of its original volume, followed by energy evaluations at each volume point. These data were fitted to the Birch–Murnaghan equation of state to determine a coarse equilibrium volume. A second, finer scan from 95\% to 105\% around this preliminary minimum was then conducted to refine the equilibrium parameters. The final equilibrium volume was used to rescale the structure without altering atomic positions, and the 0 K total energy was recomputed using the MatterSim model. This allows for volume to relax but leaves all atoms on their ideal FCC lattice points. All calculations were automated in a high-throughput workflow, with fitted E–V curves and optimized structures saved for further thermodynamic modeling. 

\subsection{Reference datasets}
To systematically assess the influence of data quantity and compositional diversity on model performance, we employed a universal machine-learning potential to generate the following reference datasets, using the decoration and relaxation approaches described in the Data generation Subsection.

(i)\textit{all-14-Multi}: 1000 entries for 14-component alloys;

(ii)\textit{all-$C_{14}^2$-Binary}: 3185 entries for binary systems, where each of the 91 binary combinations among 14 elements contains 35 data points ($35\times91=3185$);

(iii)\textit{all-13-Multi}: 1000 entries for 13-component alloys.

Based on these reference datasets, we designed a series of training subsets derived from binary and multicomponent alloy systems, as described in Sec.~\ref{sec2-1}. 

The exact method for generating \textit{Training-3} was to take 2 points from the 91 binaries in the \textit{all-$C_{14}^2$-Binary} reference dataset $n$ times, and then take $50 \times n$ points from the \textit{all-14-Multi} reference data, and then combine those to get $2 \times 91 \times n + 50 \times n$ data points.

\subsection{Redlich–Kister-based Sparse Linear Regression}
In CALPHAD modeling, the Redlich–Kister (RK) polynomial expansion is a foundational approach used to represent the excess Gibbs free energy of binary and multicomponent solutions. Specifically, the excess energy contribution of a binary system is typically modeled as:

\begin{equation}
    G^{\text{excess}}_{ij} = x_i x_j \sum_{v=0}^{n} L_{ij}^{(v)} (x_i - x_j)^v,
\end{equation}

where $x_i$ and $x_j$ are the mole fractions of the components $i$ and $j$, and $L_{ij}^{(v)}$ are the parameters of the Redlich-Kister interaction of order $v$. This form allows for a flexible yet physically grounded description of composition-dependent thermodynamic interactions.

We can fit the coefficients of this formulation with supervised regression by adopting an RK-like polynomial expansion of features. For a system with $n$ elements, we considered all unique binary interactions $(i, j)$ and defined the feature basis using up to 2nd order terms as:

\begin{equation}
    I_{ij}^{(v)} = x_i x_j (x_i - x_j)^v, \quad v = 0, 1, 2.
\end{equation}

This results in $3 \times \binom{n}{2}$ features per sample, capturing second-order interactions and asymmetries in composition dependence. These features were arranged into a design matrix and used to fit a sparse linear regression model with the Lasso algorithm, which enforces sparsity through $L_1$ regularization. The regularization parameter $\alpha$ was not fixed a priori; instead, it was re-determined in each fitting procedure via 50-fold cross-validation using the \texttt{LassoCV} routine~\cite{ranstam2018lasso} from \texttt{scikit-learn}~\cite{pedregosa2011scikit}. In practice, the cross-validation consistently selected $\alpha$ values in the range of $10^{-6}$ to $10^{-7}$, ensuring an appropriate balance between predictive accuracy and model sparsity. Note that this approach can be trivially generalized to higher-order RK terms.

\subsection{Machine learning model construction and evaluation}

To develop predictive models for formation energy, we employed the AutoGluon-Tabular framework~\cite{erickson2020autogluon}, which provides automated machine learning (AutoML) capabilities, including model selection, ensemble construction, hyperparameter tuning, and performance evaluation.
Each dataset was trained independently using AutoGluon's \texttt{high\_quality} preset, which invokes a broad suite of base models such as gradient boosting machines (LightGBM, CatBoost, bagged and randomized forests, k-nearest neighbors, neural networks, and rule-based learners (FastAI tabular model, XGBoost). These base models are trained and evaluated across multiple levels using multi-layer stack ensembling with weighted averaging, where the ensemble weights are optimized using hold-out validation scores.

During training, each model was constrained to a maximum runtime of 600 seconds, with early stopping applied whenever supported by the underlying algorithm. We verified that this budget was generally adequate by repeating selected runs with different random seeds and by extending the training limit to 1200 seconds for representative cases. These tests yielded very similar validation errors and did not materially alter the identity of the best-performing model, indicating that the 600-second budget was sufficient for convergence in the present workflow. AutoGluon then selected the best-performing model based on internal validation loss and refit it on the full training dataset before deployment.

To assess generalization performance, all trained models were evaluated against several fixed test sets withheld from the training process. The prediction quality was measured using three standard regression metrics: mean absolute error (MAE), root mean square error (RMSE), and R-squared ($R^2$). In addition to numerical scores, parity plots were generated for each model to visualize predicted versus true formation energies. The evaluation outputs were saved along with the trained predictors for reproducibility and are all available with the accompanying data and code (see Data and code availability). This automated modeling strategy enables robust and scalable benchmarking across multiple datasets while minimizing manual intervention and hyperparameter bias.

\subsection{Feature engineering strategies}

 To encode compositional information for the ML prediction of formation energies, we employed two contrasting feature engineering strategies: (i) an empirical polynomial basis inspired by Redlich-Kister (RK) expansions, and (ii) an elemental property-based descriptor set based on the Magpie framework implemented in \texttt{matminer}. These two approaches represent distinct modeling philosophies—one based on the amounts of each element and representing the classical empirical expansions used in CALPHAD thermodynamic modeling, and the other based on systematic descriptors, with physical insight provided by elemental properties. They are evaluated independently and are not used in combination.

The RK polynomial expansion~\cite{hillert1988partial} has been extensively adopted in CALPHAD assessments to represent excess Gibbs energy contributions from binary and multicomponent interactions. While rooted in the tradition of thermodynamic modeling, the RK formulation itself is empirical and does not derive from first-principles physics. In this study, we construct RK-inspired features by enumerating pairwise interaction terms over a normalized 14-element composition vector $\mathbf{x} = [x_1, x_2, \dots, x_{14}]$ corresponding to elements Mo, V, W, Si, Cu, Ti, Al, S, P, Zr, Nb, As, Sb, and Hf. The features and their fitting are described in the Results Section.

As a data-driven alternative, we employed Magpie descriptors~\cite{ward2016general, ward2018matminer} using the \texttt{ElementProperty} featurizer from the \texttt{matminer} library. These features summarize elemental properties (e.g., atomic number, atomic radius, melting point, electronegativity) using statistical aggregates across the elements present in a compound’s composition. The resulting feature vectors provide a high-quality, physically informed representation of compositional data. The final features also make no direct reference to any elements, which means that models based on the features can be expected to have some ability to extrapolate to entirely new elements.
Each compound was first parsed into a \texttt{Composition} object via \texttt{pymatgen}~\cite{ong2013python}, and the Magpie preset was applied to compute over 100 real-valued features per entry.

\backmatter

\bmhead{Acknowledgements}
We used the computational resources provided by the Center for High Throughput Computing (CHTC) at the University of Wisconsin–Madison. The authors gratefully acknowledge the computing time provided to them on the high-performance computer Lichtenberg at the NHR Centers NHR4CES at TU Darmstadt.

\bmhead{Data and code availability}
Additional supporting data are available from the corresponding authors upon reasonable request.

\bmhead{Author contributions}
C.S. performed all computations and analysis, with guidance from D.M., I.S., and M.A. All authors contributed substantially to the design of the research and the interpretation of the results. C.S. wrote the paper with input from all authors.

\bmhead{Competing Interests Statement}
The authors declare no competing interests.

\bmhead{Funding}
We gratefully acknowledge support from the Department of Energy (DOE) Office of Nuclear Energy’s (NE) Nuclear Energy University Programs (NEUP) under award number 21-24582.



\bibliography{sn-bibliography}

\end{document}